\documentclass[journal,10pt]{IEEEtran}
\usepackage{cite}
\usepackage{bbding}
\usepackage{makecell}
\usepackage{graphicx}
\usepackage{float}
\usepackage{amsthm}

\usepackage{amsmath,amssymb,amsfonts,bm,subfigure,lipsum}
\usepackage{stfloats}
\usepackage{algorithm}
\usepackage{algorithmic}
\usepackage{mathrsfs}
\usepackage{graphicx}
\usepackage[table]{xcolor}
\usepackage{threeparttable}
\usepackage{textcomp}
\usepackage{xcolor}
\usepackage{color}
\usepackage{booktabs}
\usepackage{amsmath}
\usepackage{bbm}
\usepackage{multicol}
\usepackage{multirow}
\usepackage[english]{babel}
\usepackage{upgreek}
\usepackage{float}
\usepackage{colortbl}
\usepackage[table]{xcolor}
\usepackage{cuted}
\usepackage{epstopdf}

\def\BibTeX{{\rm B\kern-.05em{\sc i\kern-.025em b}\kern-.08em
    T\kern-.1667em\lower.7ex\hbox{E}\kern-.125emX}}
\usepackage{graphicx,graphics,color,epsfig,subfigure,graphpap,rotate}
\usepackage{times, verbatim, subfigure, epsfig, graphicx, latexsym}
\usepackage{url}
\usepackage{subfigure}

\begin{document}
\title{Secure Semantic Communication \\ via  Paired Adversarial Residual
Networks}

\author{Boxiang~He,~Fanggang~Wang,~\IEEEmembership{Senior~Member,~IEEE},~and~Tony~Q.S.~Quek,~\IEEEmembership{Fellow,~IEEE}

\thanks{ Boxiang  He and Fanggang Wang are with the School of Electronic and Information Engineering, Beijing Jiaotong University, Beijing, China (e-mail: \{boxianghe1, wangfg\}@bjtu.edu.cn).

Tony Q.S. Quek  is with the  Information Systems Technology and
Design, Singapore University of Technology and Design, Singapore
(e-mail: tonyquek@sutd.edu.sg).

}

}

\maketitle


\begin{abstract}
This letter explores the positive side of the adversarial attack for the security-aware semantic communication system. Specifically, a pair of matching pluggable modules is installed:  one after the semantic transmitter and the other before the semantic receiver. The  module at transmitter uses a trainable adversarial residual network (ARN) to generate adversarial examples, while the module at receiver employs another trainable  ARN to remove the adversarial attacks and the channel noise. To mitigate the threat of semantic eavesdropping, the trainable ARNs are jointly optimized to minimize the weighted sum of the power of adversarial attack, the mean squared error of semantic communication, and the confidence of eavesdropper  correctly retrieving private information. Numerical results show that the proposed scheme  is capable of fooling the eavesdropper while maintaining the high-quality semantic communication.
\end{abstract}
\begin{IEEEkeywords}
Adversarial attack, residual network, secure semantic communication.
\end{IEEEkeywords}


\section{Introduction}

\IEEEPARstart{S}{emantic} communication  is a novel communication paradigm distinct from the traditional bit communication, aimed at conveying semantic information rather than exact bits\cite{b1,b2,b3}. With the rapid development of deep learning, various semantic communication technologies based on deep learning are widely studied and exhibit promising performance \cite{b4}. Consequently,  semantic communication has been identified by academia and industry as a key enabling technology for future sixth-generation (6G) mobile communication systems.

However, recent researches in the field of adversarial machine learning show that even a small adversarial attack, which is difficult for human vision to detect, can fool deep learning models \cite{b5}. This drives us to focus on the security of the deep-learning-enabled semantic communication systems, which face unique security threats from adversarial attacks compared to traditional communications. Research results \cite{b6} indicate that the end-to-end autoencoder communication system is more vulnerable to the adversarial attack than to the jamming attack. Currently, to resist adversarial attacks,  adversarial training is  favored by researchers because of its good performance, where the adversarial attack is added to the training samples for training\cite{b7,b8}.  Additionally, the techniques such as randomized smoothing \cite{b9} and input mask \cite{b10} also have demonstrated the unique advantages in enhancing the generalization ability to defend against the adversarial attack.

As we know, \emph{nothing interesting is ever completely onesided, so as adversarial attack.}
In our previous work\cite{b11}, by carefully designing adversarial attacks, we successfully defend against modulation eavesdropping.
In this letter,  the positive side of adversarial attack  is further explored for  the security-aware semantic communication system.  Specifically, we design a pair of matching pluggable modules via the trainable adversarial residual network (ARN)  to address the security threats of semantic communication system. The pluggable module at transmitter  utilizes a trainable ARN to produce adversarial examples, whereas the pluggable module at receiver uses a separate trainable ARN to mitigate adversarial attacks and channel noise.
The trainable ARNs are simultaneously optimized to minimize a weighted sum that includes the power of adversarial attacks, the mean squared error (MSE) of semantic communication, and the confidence level of the eavesdropper in accurately retrieving private information. Extensive experiments demonstrate that our scheme  exhibits promising performance in protecting semantic information: 1) High compatibility with existing semantic communication  systems due to the pluggable nature; 2)  low risk of privacy leakage through the use of adversarial attacks; 3) high-quality semantic communication due to  the low-power adversarial attack and the use of pluggable receiver module.

\section{Preliminaries of Adversarial Attack}
Given a deep neural network (DNN) classifier $f(\cdot;\theta)$, where $\theta$ is the parameters of DNN, the adversarial attack $\bm{\delta}$ is obtained by solving the following optimization problem:
\begin{subequations} \label{Pre_adv}
\begin{eqnarray}
&\hspace*{-3mm}\mathop{\text{max}}\limits_{\bm{\delta}}~\: p\left(f(\bm{r}+\bm{\delta};\theta)\neq f(\bm{r};\theta)\right)\\
&\hspace*{-23mm}\text{s.t.}~~ \frac{1}{N}\|\bm{\delta}\|^2\leq\epsilon
\end{eqnarray}
\end{subequations}
where $p(\cdot)$ is the probability function; $\bm{r}$ is the original input; $\epsilon$ is the power limit; $N$ denotes the dimensions of the adversarial attack $\bm{\delta}$. However, it is challenging to solve the problem \eqref{Pre_adv} directly because: i) the classifier $f(\cdot;\theta)$ and the probability function $p(\cdot)$ do not have  the explicit expressions, and ii) it is difficult to determine the convexity of the problem \eqref{Pre_adv}.

\begin{figure}[tbp]
  \centering
  \includegraphics[width=3in]{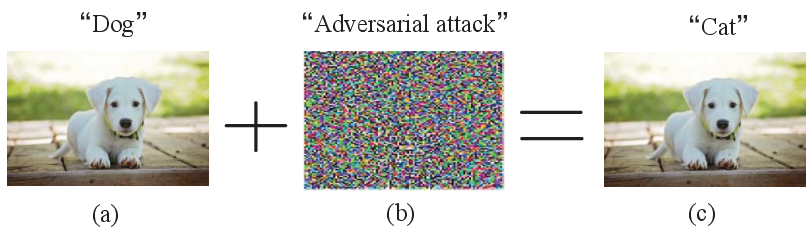}\\
  \caption{Illustrative Diagram of  the impact of adversarial attacks on
deep learning models.}  \label{adv}
\end{figure}

Alternatively, many suboptimal
methods in the computer vision domain have been proposed to generate the approximated adversarial attack using the loss function of the DNN. A common method is the Fast Gradient Sign Method (FGSM), which aims to maximize the loss function. Denote the loss function of $f(\cdot;\theta)$ as $\mathcal{L}(\bm{r}, \bm{m}, \theta)$, where $\bm{m}$ is the true label vector. The FGSM linearizes the loss function as follows
\begin{align} \label{linear}
\mathcal{L}(\bm{r}+\bm{\delta}, \bm{m}, \theta)\approx\mathcal{L}(\bm{r}, \bm{m}, \theta)+\bm{\delta}^{\mathsf{T}}\nabla_{\bm{r}}\mathcal{L}(\bm{r}, \bm{m}, \theta).
\end{align}
To maximize  \eqref{linear}, the adversarial attack is set to
\begin{align}
\bm{\delta}=\frac{\epsilon\nabla_{\bm{r}}\mathcal{L}(\bm{r}, \bm{m}, \theta)}{\|\nabla_{\bm{r}}\mathcal{L}(\bm{r}, \bm{m}, \theta)\|}.
\end{align}
To visually demonstrate the impact of adversarial attacks on deep learning models,\footnote{The evaluation of the experiments has been reported in \cite{b5}.} as shown in Fig. \ref{adv}, we provide an illustrative diagram. Fig.\ref{adv}(a) shows the original input $\bm{r}$, which the DNN classifier correctly identifies as ``Dog". Fig.\ref{adv}(b) represents the adversarial attack $\bm{\delta}$ with a small power $\epsilon$, which is added to Fig.\ref{adv}(a) to produce Fig.\ref{adv}(c). Due to the much lower power of the adversarial attack compared to the original input, human perception  recognizes Fig. \ref{adv}(c) as ``Dog". However, the DNN classifier identifies Fig. \ref{adv}(c) as ``Cat" with high confidence.


\begin{figure*}[tbp]
  \centering
  \includegraphics[width=5.3in]{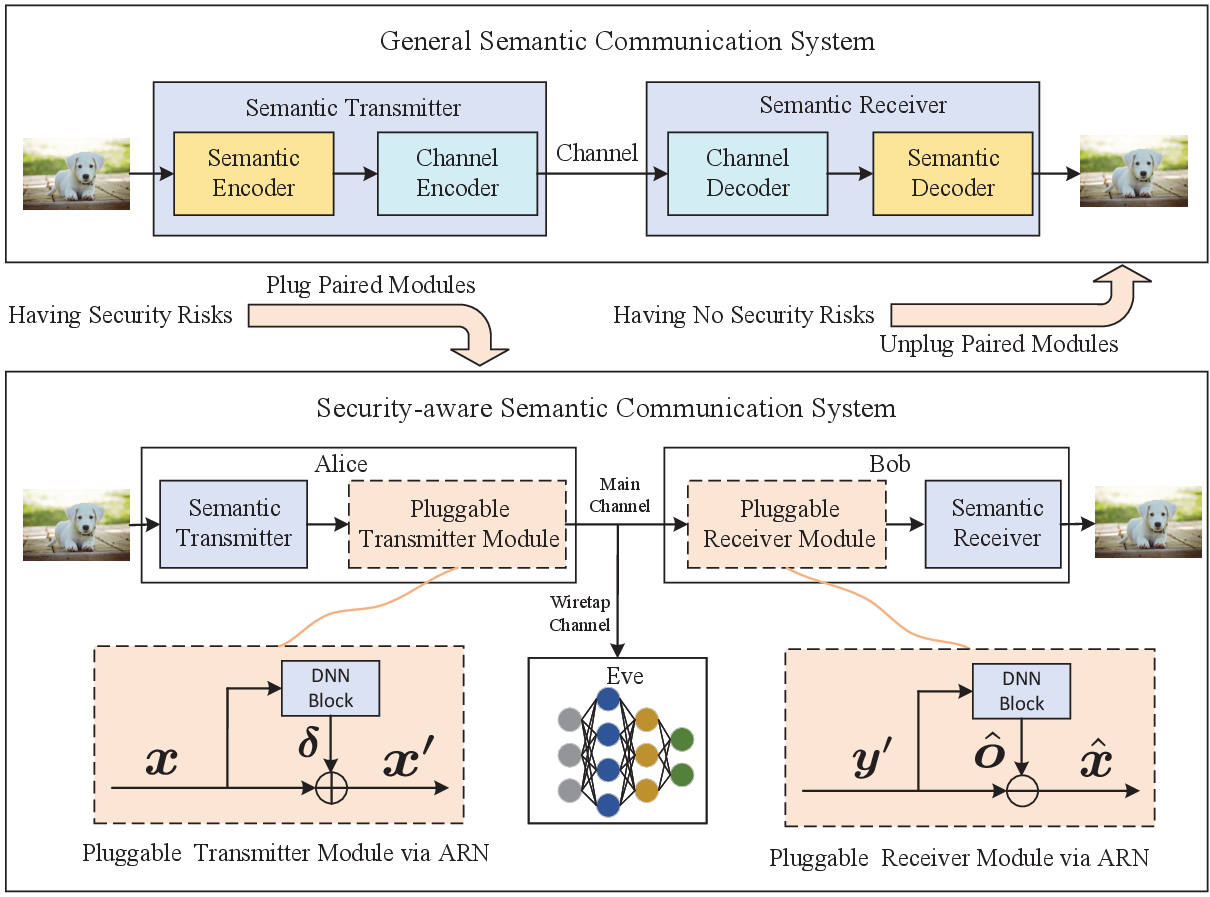}\\
  \caption{Illustrative Diagram of the general semantic communication and the proposed secure semantic communication design via paired adversarial residual networks (ARN).}  \label{CL_SC}
\end{figure*}

\section{General Semantic Communication System}
In this section, we model the general semantic communication system. As shown in the upper part of Fig. \ref{CL_SC}, the semantic encoder $\mathcal{G}(\cdot)$ first extracts the semantic information $\bm{s}'$ as
\begin{align}
\bm{s}'=\mathcal{G}\left(\bm{s};\theta_{\mathcal{G}}\right)
\end{align}
where $\bm{s}$  denotes the source data and $\theta_{\mathcal{G}}$ represents the parameters of the semantic encoder $\mathcal{G}(\cdot)$. The primary function of the semantic encoder is to extract task-relevant feature vectors. Then, the semantic information $\bm{s}'\in\mathbb{R}^{L_{s}\times1}$ is fed into the channel encoder $\tilde{\mathcal{G}}(\cdot)$ to mitigate the signal distortion caused by the wireless channel. The channel input $\bm{x}\in\mathbb{R}^{L_{x}\times1}$ is expressed by
\begin{align}
\bm{x}=\tilde{\mathcal{G}}\left(\bm{s}';\theta_{\tilde{\mathcal{G}}}\right)
\end{align}
where $L_\text{x}<L_\text{s}$ and $\theta_{\tilde{\mathcal{G}}}$ is the parameters of channel encoder $\tilde{\mathcal{G}}$. Subsequently, the signal $\bm{y}$ received at the semantic receiver can be written by
\begin{align}
\bm{y}=\bm{x}+\bm{n}
\end{align}
where $\bm{n}$ is the white noise, which follows the zero-mean Gaussian distribution with the variance $\sigma^2\bm{I}_{L_{\text{x}}}$.

Correspondingly, the output $\hat{\bm{s}}'$ of the channel decoder $\tilde{\mathcal{D}}(\cdot)$ can be denoted as
\begin{align}
\hat{\bm{s}}'=\tilde{\mathcal{D}}(\bm{y};\theta_{\tilde{\mathcal{D}}})
\end{align}
where $\theta_{\tilde{\mathcal{D}}}$ is the parameters of the channel decoder. The role of the  channel decoder is to decompress the semantic information and mitigate the effects of  the channel. Finally, the decoded signal from the semantic decoder  ${\mathcal{D}}(\cdot)$ is given by
\begin{align}
\hat{\bm{s}}={\mathcal{D}}(\hat{\bm{s}}';\theta_{{\mathcal{D}}})
\end{align}
where $\theta_{{\mathcal{D}}}$ denotes the parameters of  the semantic decoder  ${\mathcal{D}}(\cdot)$. Given a task-related loss function, such as the MSE for the data recovery  task, the  parameters set $\bm{\Theta}=\{\theta_{\mathcal{G}}, \theta_{\tilde{\mathcal{G}}}, \theta_{\tilde{\mathcal{D}}}, \theta_{{\mathcal{D}}}\}$ of the overall semantic communication system can be trained. Once the trained parameters set is obtained, the semantic encoder,  the channel encoder, the channel decoder, and semantic decoder are deployed for the practical communication task.

\section{Security-aware Semantic Communication System Design}

In this section, we first propose a security-aware semantic communication framework using the paired pluggable modules. Then, the paired pluggable modules are carefully designed via adversarial residual networks.

\subsection{Security-aware Semantic Communication Framework via Paired Pluggable Modules}

Consider a security-aware semantic communication system, as shown in the lower part of Fig. \ref{CL_SC}. Following  the naming convention in the security research, Alice, Bob, and Eve represent  the transmitter, the legitimate receiver, and the eavesdropper, respectively. To ensure that the semantic information is not obtained by eavesdroppers, Alice first feeds the output $\bm{x}$ of the semantic transmitter into a pluggable transmitter module. Then, the output of the pluggable transmitter module is written as
\begin{align}
\bm{x}'=\mathcal{E}_{\text{Alice}}(\bm{x};\theta_{\mathcal{E}_{\text{Alice}}})
\end{align}
where $\mathcal{E}_{\text{Alice}}(\cdot)$ denotes the pluggable transmitter module and $\theta_{\mathcal{E}_{\text{Alice}}}$ represents its parameters. Thus, the received signals of Bob and Eve are respectively denoted by
\begin{align}
\bm{y}'&=\bm{x}'+\bm{n}\\
\bm{y}'_{\text{Eve}}&=\bm{x}'+\bm{n}_{\text{Eve}}
\end{align}
where $\bm{n}_{\text{Eve}}$ is the zero-mean Gaussian white noise with the variance $\sigma^2_{\text{Eve}}\bm{I}_{L_{\text{x}}}$. For the legal receiver, i.e., Bob, the paired pluggable receiver module is used to remove the channel noise and recover $\bm{x}$. The output of the paired pluggable receiver module is written by
\begin{align}
\hat{\bm{x}}=\mathcal{E}_{\text{Bob}}(\bm{y}';\theta_{\mathcal{E}_{\text{Bob}}})
\end{align}
where $\mathcal{E}_{\text{Bob}}(\cdot)$ denotes the pluggable receiver module and $\theta_{\mathcal{E}_{\text{Bob}}}$ is its parameters. Finally, the decoded signal of the semantic receiver is denoted by
\begin{align}
\hat{\bm{s}}={\mathcal{D}}\left(\tilde{\mathcal{D}}(\hat{\bm{x}};\theta_{\tilde{\mathcal{D}}});\theta_{{\mathcal{D}}}\right).
\end{align}

For the eavesdropper, i.e., Eve, the deep neural network is employed to obtain the privacy information $\hat{\bm{s}}_{\text{Eve}}$, i.e.
\begin{align}
\hat{\bm{s}}_{\text{Eve}}=\mathcal{D}_{\text{Eve}}(\bm{y}'_{\text{Eve}};\theta_{\mathcal{D}_{\text{Eve}}})
\end{align}
where $\mathcal{D}_{\text{Eve}}(\cdot)$ is the deep neural network of Eve and  $\theta_{\mathcal{D}_{\text{Eve}}}$ denotes its parameters.

The design goal of the paired pluggable modules is to ensure that Bob can reliably recover the transmitted task-related information, while Eve obtains the privacy information with a high error probability. The design process of the paired pluggable modules is detailed in Section IV.B.

\emph{Remark 1:} The concept of pluggable design originates from practical considerations. Specifically, once a general semantic communication system has been trained with extensive data and deployed in a real-world system, it  becomes challenging to retrain it due to cost constraints. Furthermore,  many existing semantic communication systems developed by the current community overlook security threats. Retraining these systems when security concerns arise is deemed unacceptable in terms of cost.  Therefore, we devised paired pluggable modules: As depicted in  Fig. \ref{CL_SC}, when security is needed, we can integrate the paired modules, and when security is not required, we can remove them. The advantage is that we only need to train the paired pluggable modules without retraining the entire semantic communication system.

\subsection{Paired Pluggable Modules Design via  Adversarial Residual
Network }


Inspired by adversarial attacks discussed in  Section II, the pluggable transmitter module is designed to generate the adversarial attack to fool Eve. To avoid the need for both forward and backward passes for each input to compute the adversarial perturbation in FGSM, we use a DNN block to create adversarial examples, i.e.
\begin{align}
\bm{x}'&=\mathcal{E}_{\text{Alice}}(\bm{x};\theta_{\mathcal{E}_{\text{Alice}}})\notag\\
&=\underbrace{\bm{x}+\mathcal{F}_{\text{Alice}}(\bm{x};\theta_{\mathcal{F}_{\text{Alice}}})}_{\text{Residual Network at Alice}} \label{RES_alice}\\
&=\bm{x}+\bm{\delta}
\end{align}
where $\mathcal{F}_{\text{Alice}}(\cdot)$ represents the DNN block at Alice,  $\theta_{\mathcal{F}_{\text{Alice}}}$ is its parameters, and $\bm{\delta}$ is the generated adversarial attack by the DNN block at Alice. It is noteworthy that \eqref{RES_alice}  essentially describes a residual network, which can more readily learn the identity function because driving the residuals to zero is easier than learning to directly copy $\bm{x}$ at the network's output. For the  pluggable receiver module at Bob, the  objective is to remove the adversarial attack $\bm{\delta}$ generated at the pluggable transmitter module  and the channel noise $\bm{n}$. The output  of the pluggable receiver module at Bob is further written as
\begin{align}
\hat{\bm{x}}&=\mathcal{E}_{\text{Bob}}(\bm{y}';\theta_{\mathcal{E}_{\text{Bob}}}) \notag \\
&=\underbrace{\bm{y}'-\mathcal{F}_{\text{Bob}}(\bm{y}';\theta_{\mathcal{F}_{\text{Bob}}})}_{\text{Residual Network at Bob}} \label{RES_bob}\\
&=\bm{y}'-\hat{\bm{o}}
\end{align}
where $\hat{\bm{o}}$ is the estimation of $\bm{\delta}+\bm{n}$ (say, sum of the adversarial attack and the introduced channel noise);  $\mathcal{F}_{\text{Bob}}(\cdot)$ represents the DNN block at Bob; $\theta_{\mathcal{F}_{\text{Bob}}}$ is the  parameters of $\mathcal{F}_{\text{Bob}}(\cdot)$.

According to equations \eqref{RES_alice} and \eqref{RES_bob}, we deduce that the trainable parameters are $\theta_{\mathcal{F}_{\text{Alice}}}$ and $\theta_{\mathcal{F}_{\text{Bob}}}$. Next, we design the loss function for the security-aware semantic communication framework using
the residual network. To achieve the secure semantic communication, the designed loss function is divided into three parts: the first part characterizes the power of the adversarial attack, the second part describes the semantic communication quality of the Alice-Bob link, and the third part characterizes the privacy information leakage of the Alice-Eve link. Thus, the loss function is formulated as
\begin{align}
\mathcal{L}(\bm{s}, \bm{v}, \theta_{\mathcal{F}_{\text{Alice}}}, \theta_{\mathcal{F}_{\text{Bob}}})=\lambda_{\text{pow}}\mathcal{L}_{\text{pow}}+
\lambda_{\text{com}}\mathcal{L}_{\text{com}}+\lambda_{\text{pri}}\mathcal{L}_{\text{pri}}
\end{align}
where $\lambda_{\text{pow}}$, $\lambda_{\text{com}}$, and $\lambda_{\text{pri}}$ are the hyperparameters that control the balance of the multiple objectives and $\bm{v}$ is the image label vector of $\bm{s}$.  $\mathcal{L}_{\text{pow}}$ denotes the loss function of the power of the adversarial attack $\bm{\delta}$, which is designed as
\begin{align}
\mathcal{L}_{\text{pow}}=\text{max}\left(0, \frac{1}{N}\|\bm{\delta}\|^2-\epsilon\right)
\end{align}
where $\epsilon$ is the power limit of the adversarial attack $\bm{\delta}$. The loss function $\mathcal{L}_{\text{pow}}$ tries to heavily penalize deviations exceeding the power limit  $\epsilon$, with no penalty for using the less power. $\mathcal{L}_{\text{com}}$ is the loss function of the semantic communication quality of the Alice-Bob link, which is designed as the MSE loss, i.e.
\begin{align}
\mathcal{L}_{\text{com}}=\mathbb{E}\|\hat{\bm{s}}-\bm{s}\|^2.
\end{align}
It is worth noting that here we focus on an image data reconstruction task, so we use MSE as the loss function. However,  our approach can be extended to other tasks by modifying the loss function $\mathcal{L}_{\text{com}}$. For example, in text transmission, $\mathcal{L}_{\text{com}}$ could be replaced with text semantic similarity.  $\mathcal{L}_{\text{pri}}$ represents the loss function for the privacy leakage of the Alice-Eve link. For convenience, we consider the image label as the privacy information Eve seeks to obtain. In practice, maximizing the loss function of Eve (e.g., cross-entropy) may potentially cause the loss value to approach infinity, leading to training instability. Here, we  use the confidence of the image being classified as a true image label as the loss function, where the confidence is obtained using the softmax function. Assuming that the output of the $k$-th neuron in the final layer is used to describe the confidence that $\bm{s}$ is classified  to be the true label $\bm{v}$, the corresponding  confidence is given by
\begin{align}
z_k=\frac{e^{\varphi_k}}{\sum_{k=1}^{K}e^{\varphi_k}}
\end{align}
where $\varphi_k$ is the output of the $k$-th neuron in the final layer of Eve's classifier and $K$ represents the total number of categories. Therefore, the loss function for the privacy leakage of the Alice-Eve link is written by
\begin{align} \label{loss_pri}
\mathcal{L}_{\text{pri}}=\mathbb{E}(z_k).
\end{align}
The loss function  $\mathcal{L}_{\text{pri}}$   implies that the greater the confidence for the true class, the higher the penalty imposed.

Finally, the optimal parameters set $\{\theta_{\mathcal{F}_{\text{Alice}}}^{*}, \theta_{\mathcal{F}_{\text{Bob}}}^{*}\}$ is obtained by solving the following problem:
\begin{align} \label{opti}
 \{\theta_{\mathcal{F}_{\text{Alice}}}^{*}, \theta_{\mathcal{F}_{\text{Bob}}}^{*}\}=\mathop{\text{argmin}}\limits_{\theta_{\mathcal{F}_{\text{Alice}}}, \theta_{\mathcal{F}_{\text{Bob}}}}~\mathcal{L}(\bm{s}, \bm{v},\theta_{\mathcal{F}_{\text{Alice}}}, \theta_{\mathcal{F}_{\text{Bob}}})
\end{align}
where the problem \eqref{opti} can be solved using  stochastic gradient descent method. The training process is summarized in Algorithm $1$, where $l$ is the sample index and  $L$ represents the size of the training dataset.  From the input of  Algorithm $1$, we can observe that when training the paired pluggable modules, the parameters of the original general semantic communication system are frozen.  Thus, if there are no security requirements, we can remove the paired pluggable modules without affecting the performance of the original general semantic communication system. The benefit of this approach is that we do not need to repeatedly train the original general semantic communication system based on security requirements. Instead, we train the paired pluggable modules, which significantly reduces the training cost. Generally, semantic communication system models are large, and the cost of retraining is deemed unacceptable.

\begin{algorithm}[t]
  \label{SVM}
  \caption{Training process of the paired pluggable  modules}
   {\bf Input:}
  Training dataset $\{{\bm{s}}_{(l)},\bm{v}_{(l)}\}_{l=1}^{L}$; the trained parameters set $\bm{\Theta}^{*}=\{\theta_{\mathcal{G}}^{*}, \theta_{\tilde{\mathcal{G}}}^{*}, \theta_{\tilde{\mathcal{D}}}^{*}, \theta_{{\mathcal{D}}}^{*}\}$ of the general semantic communication system;  
  \begin{algorithmic}[1]
    \STATE{Initialize $\{\theta_{\mathcal{F}_{\text{Alice}}}, \theta_{\mathcal{F}_{\text{Bob}}}\}$;}
     \STATE{{\bfseries while} loss is not converged {\bfseries do}}
    \STATE{~~~Randomly sample ${\bm{s}}$ from training dataset;}
    \STATE{~~~~Update $\{\theta_{\mathcal{F}_{\text{Alice}}}, \theta_{\mathcal{F}_{\text{Bob}}}\}$ using  stochastic gradient descent;}
    \STATE{{\bfseries end while}}
  \end{algorithmic}
\end{algorithm}

\section{Performance Analysis}
In this section, we analyze the performance of the proposed scheme. The MNIST dataset is used to test the performance. The MNIST dataset consists of  $60000$ images, each being a $28\times28$ handwritten number. Furthermore, we utilize $50000$ images for the training and the rest $10000$ images for  testing. In our simulation,  the semantic  encoder and the semantic decoder are  two-layer fully connected neural network,  each with $64$ neurons per layer, employing  the ReLU activation function. The channel  encoder and the channel decoder are  one-layer fully connected neural network, each with $23$ neurons per layer. Since the input image size is $28\times28=784$, the compression ratio is $23/784$. The DNN block used in the paired pluggable modules is a three-layer fully connected neural network, with $23$, $64$, and $23$ neurons in each layer, respectively. It's important to note that  the number of neurons in the output layer of the DNN block should   match that of the output layer of the channel encoder/decoder. The classifier employed by Eve is  a  four-layer fully connected neural network, with $64$, $32$, $32$, and $10$ neurons in each layer, respectively. The output power of the semantic transmitter is normalized to $1$, i.e., $\mathbb{E}(\|\bm{x}\|^2)=1$. The power limit of the adversarial attack is set to $\epsilon=0.1$, which is $10\%$ of the output power of the semantic transmitter. The hyper-parameters $\lambda_{\text{pow}}$, $\lambda_{\text{com}}$, and $\lambda_{\text{pri}}$
are set to $0.1$, $0.5$, and $0.01$, respectively. Unless otherwise specified, the above settings are used for the fault parameters.

\begin{figure}[tbp]
  \centering
  \includegraphics[width=3.4in]{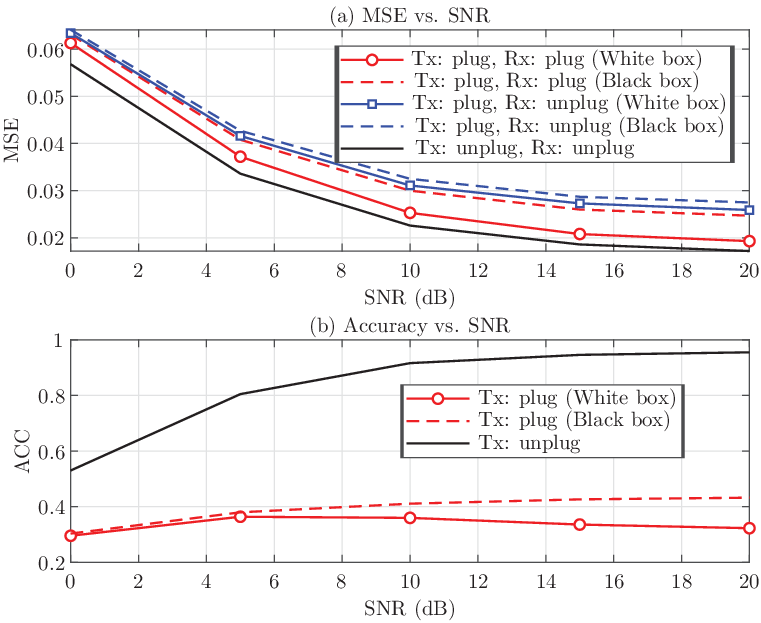}\\
  \caption{MSE and accuracy (ACC) performance evaluation w.r.t. SNR. (a) MSE vs. SNR; (b) accuracy vs. SNR. In the white box scenario, Alice is aware of the network and parameters used by Eve. Conversely, in a black box scenario, Alice lacks information available about Eve.}  \label{white_box_atn}
\end{figure}

In Fig. \ref{white_box_atn}, we evaluate the MSE of the Alice-Bob link  and the image classification accuracy  of the Alice-Eve link. ``White box" means that Alice possesses the knowledge of the structure and parameters of the deep neural network at Eve, allowing the paired modules to be trained using the true Eve's model. On the other hand,  ``Black box" indicates that Alice lacks the prior information about  Eve. In this case, we employ a local substitute model, comprising a two-layer fully connected neural network with  $16$ neurons in the first layer and $10$ neurons in the second layer.  It is observed that the local substitute model of Alice differs significantly from the true model used by Eve. First, from Fig. \ref{white_box_atn}, it is evident that
employing  the receiver pluggable module results in a better MSE performance compared to not using it. Furthermore,  by simultaneously employing the paired  pluggable modules, the Alice-Bob link  achieves a MSE performance close to that of the original semantic communication system without security consideration, while Eve's image classification accuracy drops to around $0.4$. Moreover,
 comparing the white box and black box scenarios, we note that the black box scenario only leads to  a slight decrease in MSE performance and attack capability. This finding relaxes our requirements for prior knowledge about Eve.

\begin{figure}[tbp]
  \centering
  \includegraphics[width=3.43in]{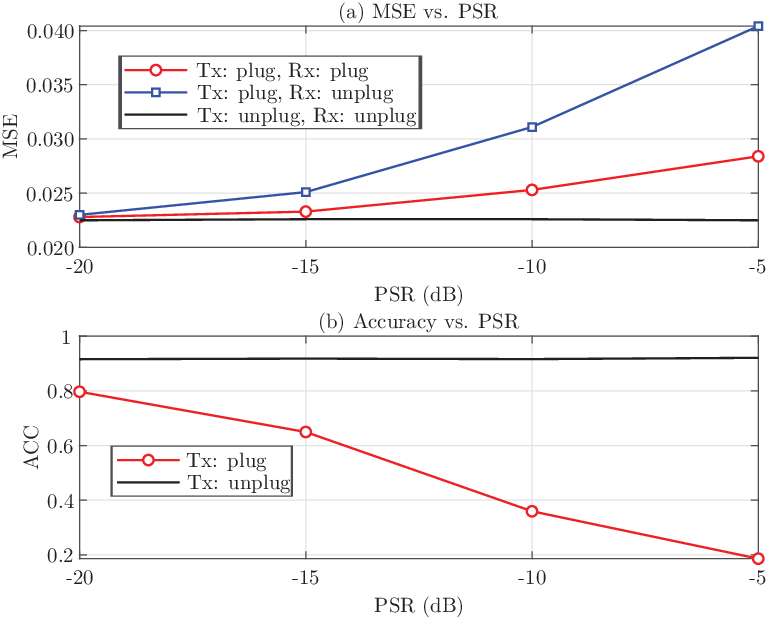}\\
  \caption{MSE and accuracy  performance evaluation w.r.t. PSR, where PSR is the power ratio of the adversarial attack and the output of the semantic transmitter. (a) MSE vs. PSR; (b) accuracy vs. PSR. }  \label{psr}
\end{figure}

In Fig. \ref{psr}, we depict the curves of MSE and classification accuracy as PSR changes, with SNR set to $10$ dB. From Fig. \ref{psr}, it is evident that even when the PSR is $-20$ dB, meaning the power limit $\epsilon$ of the adversarial attack is $1\%$ of the signal power, the adversarial attack generated by the  pluggable transmitter module  reduces Eve's classification accuracy by approximately $10\%$. As the PSR increases, we observe  that  a gradual decrease in Eve's classification accuracy. When PSR reaches $-5$ dB, Eve's classifier on the verge of failure.  Encouragingly, the MSE of the Alice-Bob link remains below $0.03$ with the use of the pluggable receiver module.

 \section{Conclusion}
 In this letter, we proposed a design scheme of the paired pluggable modules using ARN for the security-aware semantic communication system. The pluggable transmitter module using ARN was devised to generate the adversarial examples, primarily aimed at  fooling the eavesdropper. Correspondingly, the pluggable receiver  module utilizes ARN  to  eliminate  the adversarial attack and the introduced channel noise. The optimization of the paired pluggable modules is conducted carefully by considering the weighted sum of attack power, semantic communication quality, and privacy information leakage.
Simulation results demonstrate the excellent performance of the proposed scheme in the security-aware semantic communication system.

 This letter presented  initial results on ARN enabling secure semantic communication, demonstrating  promising performance even with the  fully connected networks used as DNN blocks of ARN. However, it is anticipated that by replacing DNN blocks with advanced deep models such as transformers, better performance can be achieved, which will be a focus of future research. Additionally, evaluating performance on more datasets is also an important area for future work.


\begin{thebibliography}{1}
\bibitem{b1} Z. Qin, X. Tao, J. Lu, W. Tong, and G. Ye Li, ``Semantic communications: Principles and challenges," \emph{arXiv:2201.01389,} 2021.
\bibitem{b2} W. Yang et al., ``Semantic communications for future internet: Fundamentals, applications, and challenges," \emph{IEEE Commun. Surveys Tuts.,} vol. 25, no. 1, pp. 213-250, 1st Quart. 2023.
\bibitem{b3} X. Luo, H.-H. Chen, and Q. Guo, ``Semantic communications: Overview open issues and future research directions," \emph{IEEE Wireless Commun.,} vol. 29, no. 1, pp. 210-219, Feb. 2022.
\bibitem{b4} H. Xie, Z. Qin, G. Y. Li, and B.-H. Juang, ``Deep learning enabled semantic communication systems," \emph{IEEE Trans. Signal Process.,} vol. 69, pp. 2663-2675, 2021.
\bibitem{b5} A. Madry, A. Makelov, L. Schmidt, D. Tsipras, and A. Vladu, ``Towards deep learning models resistant to adversarial attacks," in \emph{Proc. Int. Conf. Learn. Represent.,} May 2018, pp. 1-10.
\bibitem{b6} M. Sadeghi and E. G. Larsson, ``Physical adversarial attacks against end-to-end autoencoder communication systems," \emph{IEEE Commun. Lett.,} vol. 23, no. 5, pp. 847-850, May 2019.
\bibitem{b7} Q. Hu, G. Zhang, Z. Qin, Y. Cai, G. Yu, and G. Y. Li, ``Robust semantic communications against semantic noise,'' in \emph{Proc.  IEEE 96th Veh. Technol. Conf. (VTC-Fall),} Sep. 2022, pp. 1-6.
\bibitem{b8} G. Nan et al., ``Physical-layer adversarial robustness for deep learning-based semantic communications," \emph{IEEE J. Sel. Areas Commun.,} vol. 41, no. 8, pp. 2592-2608.
\bibitem{b9} B. Kim, Y. E. Sagduyu, K. Davaslioglu, T. Erpek, and S. Ulukus,
``Channel-aware adversarial attacks against deep learning-based wireless
signal classifiers," \emph{IEEE Trans. Wireless Commun.,} vol. 21, no. 6, pp.
3868-3880, Jun. 2022.
\bibitem{b10} Q. Hu, G. Zhang, Z. Qin, Y. Cai, G. Yu, and G. Y. Li, ``Robust semantic communications with masked VQ-VAE enabled codebook," \emph{IEEE Trans. Wireless Commun.,} vol. 22, no. 12, pp. 8707-8722, Dec. 2023.
\bibitem{b11} B. He and F. Wang, ``Anti-modulation-classification transmitter design against deep learning approaches," \emph{IEEE Trans. Wireless Commun.,} doi: 10.1109/TWC.2023.3335050.
\end{thebibliography}
\end{document}